\documentclass[11pt]{article}
\usepackage[centertags]{amsmath}
\usepackage[latin1]{inputenc}
\usepackage{amssymb,amsthm}
\usepackage{amsmath}
\usepackage{amsfonts}
\usepackage{textcomp}
\usepackage{newlfont}
\usepackage{graphicx}
\usepackage{float}
\usepackage{eurosym}
\usepackage[dvips]{color}

\newtheorem{theorem}{Theorem}[section]
\newtheorem{proposition}{Proposition}[section]

\newtheorem{rem}{Remark}[section]

\newtheorem{ex}{Example}[section]
\newcommand{\e}{\mathrm{e}}
\newcommand{\E}{\mathbb{E}}

\newcommand{\ci}{\mathrm{i}}

\def\P{{\mathbb{P}}}

\def\half{\frac{1}{2}}
\def\Var{\mathrm{VaR}}
\def\CVar{\mathrm{CVaR}}

\title{On a Transform Method for the Efficient Computation of Conditional VaR
(and VaR) with Application to Loss Models with Jumps and Stochastic Volatility}
\author{Alessandro Ramponi \\ \\
Department of Economics and Finance \\
University of Roma - Tor Vergata\\
via Columbia, 2 - 00133 Roma, Italy \\
e-mail: alessandro.ramponi@uniroma2.it}

\date{}

\begin{document}

\maketitle

\begin{abstract}

In this paper we consider Fourier transform techniques to efficiently compute the Value-at-Risk and the Conditional Value-at-Risk of an arbitrary loss random variable, characterized by having a computable generalized characteristic function. We exploit the property of these risk measures of being the solution of an elementary optimization problem of convex type in one dimension for which Fast and Fractional Fourier transform can be implemented. An application to univariate loss models driven by L\'{e}vy or stochastic volatility risk factors dynamic is finally reported.

\medskip




\end{abstract}

\section{Introduction}

Measuring risks is certainly one of the core competence of any financial institution, even from a regulatory perspective. An efficient and reliable computation of risk measures is consequently a primary concern of any modern risk management activity, dramatically highlighted during the recent financial crisis. Value-at-Risk and Conditional (or Average) Value-at-Risk (henceforth VaR and CVaR) are without doubt among the best known monetary risk measures. Since its introduction, VaR rapidly has become a benchmark in the financial industry both for regulatory purposes and in the practice of risk management, mainly due to its simplicity. On the other side, it is sharply criticized for the lack of sub-additivity and the inability to quantify the severity of an exposure to rare events. For these reasons, alternative risk measures are considered as the CVaR, which turns out to be one of the well known example of coherent risk measure (see \cite{AT02},\cite{ADEH}).

In this paper we present a method for their computation based on a Fourier transform technique. The Fourier representation of distribution functions and expected values of random variables is well-known in the financial and actuarial literature: L\`{e}vy and Gil-Pelaez inversion formulas and more recently the method introduced by Carr and Madan in their seminal paper \cite{Carr99} for contingent claim valuation through Fast Fourier transform, nowadays are standard computational techniques to efficiently solve statistical and pricing problems. Here we consider a parametric framework, thus assuming a probabilistic description for the quantity we are interested in, namely profit/loss random variables. Aside the standard definitions for VaR and CVaR, in this paper we exploit an alternative  characterizations of these risk measures in which transform representation can play an important tool for their efficient calculation. In particular, their property of being the solution of an elementary optimization problem of convex type in one dimension permits to evaluate both measures by solving numerically a unique simple univariate minimization problem, in which the objective function can be efficiently computed by means of Fourier representation. A quick and dirty solution is then available through (fractional) Fast Fourier Transform algorithms.

Analytical calculation of VaR by using inversion formulas has been considered in Duffie and Pan \cite{DP} while the use of generalized Fourier transform and the FFT algorithm is more recent, see e.g. Le Courtois and Walter \cite{LCW} who applied such a technique in a Variance Gamma model, Kim et al. \cite{KRBF}, Scherer et al. \cite{SRKF} where the class of tempered stable and infinitely divisible distribution were considered or Bormetti et al. \cite{Bor10} for an application to a stochastic volatility model. The techniques considered here can be applied to all models having an analytic (and computable) characteristic function. This is the case of many financial dynamic models of returns emerged in the literature of the last decades for the analysis and the management of portfolio risks, such as L\`{e}vy finite/infinite activity and stochastic volatility models; an application to an univariate loss model in such a framework will be presented in the numerical section. But they can also be applied to that stochastic models which find their application in actuarial science, such as the compound Poisson loss distribution, used to model the aggregate claims of an insurance-risk business (see \cite{Dufresne09}).

The paper is organized as follows: VaR and CVaR are briefly introduced in Section 2, where their main characterizations are outlined. Transform technique is reviewed in Section 3 together with the use of fast and fractional Fourier transform algorithms and finally a set of numerical experiment is reported in Section 4 to show the effectiveness of the proposed procedures. In particular, the Fourier based techniques are applied to a univariate loss model of portfolio returns characterized by dynamical risk factors with jumps and stochastic volatility.

\section{VaR and Conditional VaR}

Let $(\Omega, F, \P)$ be a probability space, $L$ a real-valued random variable and $F_L(x) = \P(L \leq x)$ its distribution function. In the following we suppose that $\E[|L|] < \infty$. To measure the risk of a financial position characterized by an uncertain future value over a given time horizon, the quantiles of its distribution function are commonly used. Given a confidence level $\alpha \in (0,1)$, the set of $\alpha$-quantiles of the random variable $L$ is the interval $[q^-_{\alpha}(L), q^+_{\alpha}(L)]$ where
$$
q^-_{\alpha}(L) = \inf \{q \in \mathbb{R} | P(L\leq q) \geq \alpha \}, \ \ q^+_{\alpha}(L) = \inf \{q \in \mathbb{R} | P(L\leq q) > \alpha \}.
$$

In this paper the random variable $L$ describes the loss of a financial position. Given $L$,  VaR is defined as the lower $\alpha$-quantile, $q^-_{\alpha}(L)$:
$$
\Var_{\alpha}(L) := \inf \{q \in \mathbb{R} | P(L \leq q) \geq \alpha \}.
$$

In financial terms, VaR is ``the maximum possible loss which is not exceeded with probability $\alpha$", or ``the smallest amount of capital which, if added to the current position, keeps the probability of a non-negative outcome below the level $1-\alpha$". For a random variable having continuous and strictly increasing distributions function, $q^-_{\alpha}(L) = q^+_{\alpha}(L) \equiv q_{\alpha}(L) = F^{-1}_L(\alpha)$, the ordinary inverse of $F$, i.e. VaR solves the equation
$$
\P( L \leq \Var_{\alpha}(L)) = \alpha \ \ \ \ \ (\mbox{or equivalently} \ \ \P( L \geq \Var_{\alpha}(L)) = 1-\alpha).
$$

Although widely used, it is well known that VaR is not a coherent risk measure (see \cite{AT02}, \cite{ADEH}), in particular for being not sub-additive. To overcome the weakness of VaR, several alternative risk measures have been proposed in literature, among which the Conditional, or Average, Value-at-Risk, which does satisfy the axioms of coherence (see e.g.  \cite{AT02},\cite{ADEH}). Several equivalent definitions have been proposed in the literature: given the confidence level $\alpha \in (0,1)$, the basic idea is to average all the possible losses exceeding $\Var_{\alpha}(L)$, that is
$$
\CVar_{\alpha}(L) := \frac{1}{1-\alpha} \int_{\alpha}^1 \Var_{u}(L) \ du.
$$

Alternatively, it may be convenient to define the CVaR as the expectation of $L$ under the (scaled) distribution function (see \cite{RU02})
$$
F_{L,\alpha}(x) = \left \{ \begin{array}{cc}
               0 & \mbox{for} \ \ x < q_{\alpha}^-(L)\\
               (F_L(x)-\alpha) / (1-\alpha) & \mbox{for} \ \ x \geq q_{\alpha}^-(L). \\
               \end{array} \right.
$$

\bigskip

When $F$ is continuous and strictly increasing then $\Var_{\alpha}(L) = F^{-1}_L(\alpha)$ and
$$
\int_{\alpha}^1 \Var_{u}(L) \ du = \int_{F^{-1}_L(\alpha)}^{+\infty} x \ dF_{L}(x) = \E[X \mathbb{I}_{X \geq F^{-1}_L(\alpha)}];
$$
therefore $\CVar_{\alpha}(L) = \E[L | L \geq \Var_{\alpha}(L)]$\footnote{For general distribution functions this is not true: see \cite{RU02} and \cite{Hur02} for a detailed discussion about alternative definitions.}. Furthermore, since $x I_{x \geq q} = (x-q)^+ + q I_{x \geq q}$ we get
\begin{equation} \label{avar0} \CVar_{\alpha}(L) = \Var_{\alpha}(L) + \frac{1}{1-\alpha} \E[(X-\Var_{\alpha}(L))^+].
\end{equation}

More generally, it is possible to prove that (\ref{avar0}) is still valid for any $\alpha$-quantile of $L$: that is
$$
\CVar_{\alpha}(L) = q + \frac{1}{1-\alpha} \E[(L-q)^+]
$$
for any $q \in [q^-_{\alpha}(L), q^+_{\alpha}(L)]$, which evaluated for $q = \Var_{\alpha}(L)$, clearly gives (\ref{avar0}). The quantity $SL_{L}(q) \equiv \E[(L-q)^+]$ is known as the \emph{stop-loss} transform of $L$. The following approach can therefore be considered for computing VaR and CVaR:

\begin{center}
\fbox{
\begin{minipage}{15cm}
\textsc{Algorithm 1. - \textsc{Two Steps}}

\begin{enumerate}
        \item compute $\Var_{\alpha}(L)$, i.e. find $q^*$ such that
                     $\P(L \leq q^*) = \alpha$ (possibly not a unique solution);
        \item compute $\E[(L-q^*)^+]$ to get $\CVar$ through formula (\ref{avar0}).
\end{enumerate}
\end{minipage}
}
\end{center}

\bigskip

An alternative characterization of VaR and CVaR for an arbitrary loss $L$, due to Rockafellar and Uryasev \cite{RU00}, \cite{RU02}, is obtained as follows. For a given value $\alpha \in (0,1)$ let us introduce the real function
$$
G_{L,\alpha}(x) = x + \frac{1}{1-\alpha} \E[(L-x)^+], \ \ x \in \mathbb{R}
$$
and let $\Gamma = \mathrm{argmin}_x G_{L,\alpha}(x)$ be the set of $x$ for which the minimum value of $G_{L,\alpha}$ is attained, with $\Gamma^-$ and $\Gamma^+$ respectively equal to the lower and the upper endpoint of $\Gamma$. The proof of the following Theorem can be found in \cite{RU02} (but see also \cite{FS}, where an alternative proof based on the Fenchel-Legendre transform is proposed (Lemma 4.6)):

{\theorem \label{theo1} For any random variable $L$, the function $G_{L,\alpha}(\cdot)$ is finite and convex with
$$
\CVar_{\alpha}(L) = \min_{x \in \mathbb{R}} G_{L,\alpha}(x).
$$

Moreover $\Gamma$ is a nonempty, closed, bounded interval with $q_{\alpha}^-(L) = \Gamma^-$ and $q_{\alpha}^+(L) = \Gamma^+$. In particular, one always has
$$
q_{\alpha}(L) \in \Gamma, \ \ \CVar_{\alpha}(L) = G_{L,\alpha}(q_{\alpha}(L)).
$$
}

\bigskip

This theorem suggests to compute VaR and CVaR by solving an unique optimization problem:

\begin{center}
\fbox{
\begin{minipage}{15cm}
\textsc{Algorithm 2. - \textsc{NL-Min}}
\begin{enumerate}
       \item Solve the non-linear minimization problem $\min_x  G_{L,\alpha}(x)$.
\end{enumerate}

\end{minipage}
}
\end{center}

\begin{figure}[t]
\begin{center}
\includegraphics[width=16cm,height=11cm]{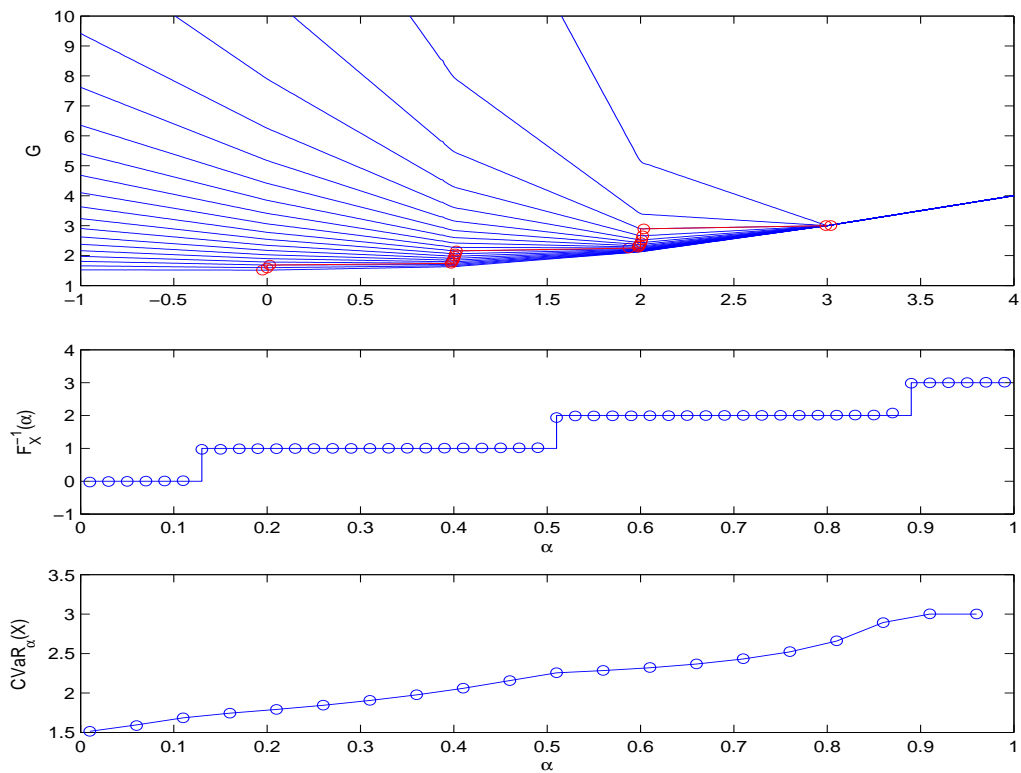}
\caption{\small  Binomial model, $X \sim Binomial(N,p)$. In the
upper plot, the functions $G$ and the corresponding solutions of
the minimization problem, for $\alpha \in [0.01,
0.99]$. In the lower plots, true VaR and CVaR (continuous lines)
and the corresponding values ('o') computed by solving the
minimization problem.} \label{Fig1}
\end{center}
\end{figure}

\bigskip

In both the algorithms,  the expectations required to compute VaR and CVaR must be evaluated for different values of a parameter $q$, namely $\P(L \leq q) = \E[\mathbb{I}_{\{L \leq q \}}]$ and $\E[(L-q)^+]$. This can be efficiently done by means of the Fourier transform technique. Before recalling the basic properties of this class of computational methods, we now briefly introduce a simple loss model which will be used to test the computational procedures in the final Section.

\begin{figure}[h]
\begin{center}
\includegraphics[width=16cm,height=11cm]{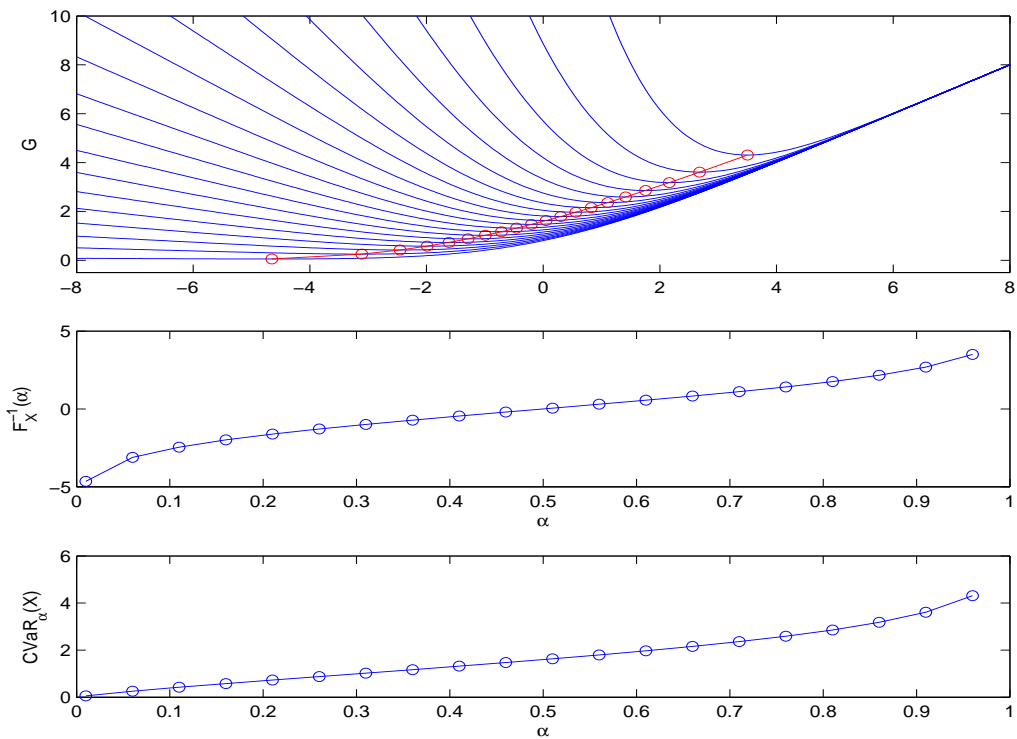}
\caption{\small  Gaussian model, $X \sim N(\mu,\sigma^2)$. In the
upper plot, the functions $G$ and the corresponding solutions of
the minimization problem, for $\alpha \in [0.01,
0.99]$. In the other plots, true VaR and CVaR (continuous lines)
and the corresponding values ('o') computed by solving the
minimization problem.} \label{Fig2}
\end{center}
\end{figure}

\paragraph{An Univariate Loss Model.} Let us consider on a filtered probability space $(\Omega, \mathcal{F}, \mathcal{F}_t, \P)$ a stochastic process of the form $V_t=V_0\e^{X_{t}}$, $V_0 >0$, modeling the value of a risky position for $t \in [0,T]$. The random variable we consider is
$$
L=V_0 \e^{rT}-V_T =V_0 \e^{r T} - V_0 \e^{X_T}
$$
where $r$ is the risk-free interest rate, that we can assume as a deterministic constant in the reference period for easiness of notation. In such a case the function $G_{L,\alpha}$ becomes
\begin{eqnarray}
G_{L,\alpha} \equiv G^{(e)}_{L,\alpha}(x) & = & x + \frac{1}{1-\alpha} \E[(V_0 \e^{r T} - V_0 \e^{X_T}-x)^+] \nonumber \\
& = & V_0 \left( \frac{x}{V_0} + \frac{1}{1-\alpha} \E \left[\left(\left(\e^{r T}-\frac{x}{V_0}\right)- \e^{X_T}\right)^+\right] \right) \nonumber \\
& = & \left \{
\begin{array}{lc}
             x & x \geq V_0 \e^{rT} \\ \\
             V_0 \left( \frac{x}{V_0} + \frac{1}{1-\alpha}
             \E \left[\left(\e^{k(x/V_0)} -
             \e^{X_T}\right)^+\right] \right) & x < V_0 \e^{rT}
           \end{array} \right. \label{GfunEx}
\end{eqnarray}
where $k(v) = \log(\e^{rT}-v)$. Theorem (\ref{theo1}) still applies, with the constraint $x < V_0 \e^{rT}$.

\bigskip

{\ex \label{CvarBS} Let us consider the classical log-normal model, where $X_T=(\mu-\sigma^2/2)T+\sigma W_T$, $W_t$ being the Wiener process, $\mu \in \mathbb{R}$ and $\sigma >0$ two given parameters. In such a case, standard calculations yield
$$
VaR_{\alpha}(X) = V_0 \e^{rT} - V_0 \e^{(\mu-\sigma^2/2)T+\sigma \sqrt{T} z_{1-\alpha}}
$$
and
$$
G_{\alpha}(x) = x +\frac{1}{1-\alpha}[(V_0 \e^{rT}-x) N[-d_2(x)] - V_0 \e^{\mu T} N[-d_1(x)]]
$$
\begin{eqnarray*}
d_1(x) & = & \frac{1}{\sigma \sqrt{T}}\left (\log \left(\frac{V_0}{V_0 \e^{rT}-x} \right) +(\mu + \sigma^2/2) T\right), \\
d_2(x) & = & \frac{1}{\sigma \sqrt{T}}\left (\log \left(\frac{V_0}{V_0 \e^{rT}-x} \right) +(\mu - \sigma^2/2)
T\right)
\end{eqnarray*}
where $N[d]$ is the standard normal cumulative distribution function and $z_{1-\alpha}$ is the corresponding
$(1-\alpha)$-quantile (or critical value). Finally,
$$
CVaR_{\alpha}(X) = V_0 (\e^{rT} - \e^{(\mu-\sigma^2/2)T+\sigma  \sqrt{T} z_{1-\alpha}}) + \frac{V_0 \e^{\mu T}}{1-\alpha} \left( \e^{-\sigma^2 T/2 + \sigma \sqrt{T} z_{1-\alpha}} N[z_{1-\alpha}] - N[z_{1-\alpha}-\sigma \sqrt{T}] \right).
$$
}

\section{The Fourier transform method}

Fourier transform methods are efficient techniques emerged in recent years in the financial practice as one of the main methodology for the evaluation of derivatives. In fact, the no-arbitrage price of an European style contingent claim can be represented as the (conditional) expectation of the derivative payoff under a proper risk-neutral measure (see e.g. \cite{Bjork}). These methods essentially consist on the representation of such an expectation as the \emph{convolution} of two Fourier transforms. Since the value of most derivatives depends on a trigger parameter, two main variants have been developed depending on which variable of the payoff is transformed into the Fourier space. In our setting, due to the functional (exponential) form of the transformed functions we consider, the formulas we get by applying the two approaches are essentially the same: we can pass from one to the other by simply changing the integration contour (see \cite{Ram10}). In the following we choose to work with the generalized Fourier transform (GFT) w.r.t. the trigger parameter $v$. In essence, given a function $H(y,v)$, the quantity that we want to compute is $h(v) = \E[H(Y,v)]$, where the expectation is taken over a given probability measure $\P$. Let us consider the GFT with respect to $v$: formally we have for $z = u + \ci \nu \in \mathcal{C} \subset \mathbb{C}$
$$
\hat h(z) = \int_{\mathbb{R}} \e^{\ci z v} h(v) dv =  \int_{\mathbb{R}} \e^{\ci z y} \left( \int_{\mathbb{R}} H(y,v) \mathbb{P}(dy)\right) dv = \int_{\mathbb{R}} \widehat H^{(v)}(z,y)  \mathbb{P}(dy) = \E[\widehat H^{(v)}(z,Y)]
$$
where we have defined
$$
\widehat H^{(v)}(z,y) = \int_{\mathbb{R}} \e^{\ci z v} H(y,v) dv.
$$
Notice that $\hat h$ corresponds to the classical Fourier transform of the $\nu$-damped expectation, as introduced in \cite{Carr99}: Fourier inversion gives
$$
h(v) = \frac{1}{2\pi} \int_{\ci \nu - \infty}^{\ci \nu +\infty} \e^{-\ci z v} \widehat h(z) dz = \frac{1}{2\pi} \int_{\ci \nu - \infty}^{\ci \nu +\infty} \e^{-\ci z v} \E[\widehat H^{(v)}(z,Y)] \ dz
$$
for $\nu$ in some strip of $\mathbb{C}$. The previous equalities must be justified under the appropriate conditions on the function $H$, its transform and the characteristic function of the underlying random variables (see e.g. \cite{Lee04} for a thorough discussion on the subject). For our application we consider the following functions:
$$
H_1(y,v) = (y-v)^+, \ \ \ H_2(y,v) = \mathbb{I}_{\{y \leq v\}},
$$
$$
H_3(x,k) = (e^x-e^k)^+, \ \ \ {\bar H}_3(x,k) = (e^k-e^x)^+.
$$

The reason for considering an exponential transformation of the basic risk factor is the fact that many financial models are usually introduced in the form $\exp(X)$, as in Example (\ref{CvarBS}). Their GFT are readily obtained by means of standard (complex) integration: we summarize such results in the following

{\proposition Let $z = u + \ci \nu \in \mathbb{C}$, then
\begin{equation}
\widehat{H}_1^{(v)}(z,y)  =  \frac{\e^{(\ci u -\nu)y}}{(\ci u -
\nu)^2} = - \frac{\e^{\ci z y}}{z^2}, \ \ \nu < 0
\end{equation}

\begin{equation}
\widehat{H}_2^{(v)}(z,y) = \frac{\e^{(\ci u - \nu) y}}{\ci u -
\nu} = - \frac{\ci}{z} \e^{\ci z y}, \ \ \nu > 0
\end{equation}

\begin{equation}
\widehat{H}_3^{(k)}(z,x)  =  \frac{\e^{(\ci u -\nu +1)x}}{(\ci u
- \nu)(\ci u - \nu +1)} = \frac{\e^{\ci(z-\ci) x}}{\ci z - z^2}, \
\ \ \nu < 0,
\end{equation}
\begin{equation}
\widehat{\bar{H}}_3^{(k)}(z,x) =  \frac{\e^{(\ci u -\nu
+1)x}}{(\ci u - \nu)(\ci u - \nu +1)} = \frac{\e^{\ci(z -\ci)
x}}{\ci z - z^2}, \ \ \ \nu > 1.
\end{equation}
}

\bigskip

Let $\phi_Y(z) = \E[\e^{\ci z Y}], \ \ \ z \in \mathbb{C}$ be the (generalized) characteristic function of the r.v. $Y$. The following result can be proved as in \cite{Lee04}.

{\theorem

(a) If $\E[\e^{-\nu Y}] < \infty$, $\nu < 0$, then
\begin{equation} \label{H1}
H_1 = \E[(Y-v)^+] = - \frac{1}{2\pi} \int_{\ci \nu - \infty}^{\ci \nu +\infty} \e^{-\ci
z v} \frac{\phi_Y(z)}{z^2} \ dz = - \frac{1}{\pi} \int_{\ci \nu - 0}^{\ci \nu +\infty} \Re \left \{ \e^{-\ci z v} \frac{\phi_Y(z)}{z^2} \right \}  dz;
\end{equation}

\noindent (b) Let $\bar H_2 = \frac{1}{2}(\P(Y\leq v) + \P(Y < v))$. If $\E[\e^{-\nu Y}]< \infty$, $\nu > 0$, then
\begin{equation} \label{H2}
\bar H_2 = \frac{1}{2\pi} \lim_{M \rightarrow + \infty}\int_{\ci \nu - M}^{\ci \nu +M}
\e^{-\ci z v} \frac{\ci}{z} \phi_Y(z) \ dz = \frac{1}{\pi} \int_{\ci \nu - 0}^{\ci \nu +\infty}
\Re \left\{ \e^{-\ci z v} \frac{\ci}{z} \phi_Y(z) \right\} dz;
\end{equation}

\noindent (c) If $\E[\e^{(-\nu +1) X}]< \infty$, $\nu < 0$, then

\begin{equation} \label{H3}
H_3 = \E[(\e^X-e^k)^+] = \frac{1}{2\pi} \int_{\ci \nu - \infty}^{\ci \nu +\infty} \e^{-\ci z k} \frac{\phi_X(z-\ci)}{\ci z - z^2} \ dz = \frac{1}{\pi} \int_{\ci \nu - 0}^{\ci \nu +\infty} \Re \left\{ \e^{-\ci z k} \frac{\phi_X(z-\ci)}{\ci z - z^2} \right\}  dz;
\end{equation}

\noindent (d) If $\E[\e^{(-\nu +1) X}]< \infty$, $\nu > 1$, then

\begin{equation} \label{bH3}
\bar H_3 = \E[(e^k - \e^X)^+] = \frac{1}{2\pi} \int_{\ci \nu - \infty}^{\ci \nu +\infty} \e^{-\ci z k} \frac{\phi_X(z-\ci)}{\ci z - z^2} \ dz = \frac{1}{\pi} \int_{\ci \nu - 0}^{\ci \nu +\infty} \Re \left\{ \e^{-\ci z k} \frac{\phi_X(z-\ci)}{\ci z - z^2} \right\} dz.
\end{equation}
}

\bigskip

{\rem It is worth noting that formula (\ref{H2}) is a slight generalizations of the well-known Levy's Inversion (and Gil-Pelaez) formulas, that can be obtained by using the Residue Theorem. Furthermore, in the framework of the univariate loss model discussed in Example (\ref{CvarBS}), we clearly have $\P(L \leq q) = 1-\P(\e^{X_T} < \e^k)$, where $k=\log((V_0\e^{rT}-q)/V_0)$ for $q < V_0 \e^{rT}$, and we notice that the probability of the events $\{e^X \leq e^k \}$ has the same integral representation as (\ref{H2}).
}

\bigskip

Under the hypothesis of Theorem (\ref{theo1}) we finally obtain an integral representation for the function $G_{L,\alpha}$:
\begin{eqnarray}
G_{L,\alpha}(x) & = & x - \frac{1}{(1-\alpha)2 \pi} \int_{\ci \nu -
\infty}^{\ci \nu +\infty} \e^{-\ci z x} \frac{\phi_Y(z)}{z^2} \ dz
\nonumber \\
 & = & x - \frac{\e^{\nu x}}{(1-\alpha) \pi}
\int_{0}^{+\infty} \mathrm{Re} \left(\e^{-\ci u x}
\frac{\phi_Y(u+\ci \nu)}{(u+\ci \nu)^2}\right) \ du, \ \ \ \nu < 0
\label{Gfun0}
\end{eqnarray}
or, in the case of exponential models,
$$
G^{(e)}_{L,\alpha}(x)  =  x + \frac{1}{(1-\alpha) 2\pi} \int_{\ci \nu -
\infty}^{\ci \nu +\infty} \e^{-\ci z x} \frac{\phi_X(z-\ci)}{\ci z
-z^2} \ dz
$$
\begin{eqnarray}
 =  x + \frac{\e^{\nu x}}{(1-\alpha) \pi}
\int_{0}^{+\infty} \mathrm{Re} \left(\e^{-\ci u x}
\frac{\phi_X(u+\ci(\nu-1))}{\nu^2-\nu-u^2+\ci u (1-2 \nu)}\right)
\ du, \ \ \nu > 1, \ \mbox{or} \ \ \nu <0. \label{Gfun1}
\end{eqnarray}

As it is widely known, the transform method deserves for an efficient evaluation of expectations by means of the FFT algorithm for a proper range of the trigger parameter. Actually, if only one value has to be evaluated for a fixed parameter $v$ or $k$, there is no need to use FFT and a proper quadrature algorithm suffices to compute the required expectations.

\paragraph{Fast Fourier Transform - FFT.} This technique involves two steps:
\begin{itemize}
\item a numerical quadrature scheme to approximate through a $N$-point sum the integral
$$
I(x)=\frac{1}{\pi} \int_{0}^{+\infty}  \Re \left[ \e^{-\ci u x} F(u)\right] du.
$$

By using an equi-spaced grid $\{ u_n \}_{n=1,\ldots,N}$  of the line $\{z=u + \ci v \in \mathbb{C}: u\in \mathbb{R}^+, v=\nu \}$ with spacing $\Delta$, we have
$$
I(x) \approx \Sigma_N(x) = \frac{\Delta}{\pi}\sum_{n=0}^{N-1} \Re \left [ \e^{-\ci u_n x} F(u_n) w_n \right],
$$
where $w_n$ are the integration weights\footnote{Different spacing rules can be implemented, e.g. the midpoint rule.};
\item given a grid $x_m = x_1 + \gamma m$, $m=0,\ldots N-1$, denoted by $\mathbf{x}$, the sum $\Sigma_N(x_m)$ is written as a
discrete Fourier transform (DFT) when
\begin{equation} \label{fft_vincolo}
\Delta \cdot \gamma = \frac{2 \pi}{N}
\end{equation}
that is
$$
\Sigma_N(x_m) = \frac{\Delta}{\pi} \sum_{n=0}^{N-1} \e^{- \ci n m \Delta \gamma} \e^{-\ci n \Delta x_1} F(n \Delta) w_n = \frac{\Delta}{\pi} \sum_{n=0}^{N-1} \e^{- \ci n m \frac{2 \pi}{N}} h_n
$$
where
\begin{equation} \label{hvec}
h_n = \e^{-\ci n \Delta x_1} F(n \Delta) w_n.
\end{equation}
\end{itemize}

The integral $I(x)$ is therefore approximated over the grid $\mathbf{x}$ as $I(x_m) \approx \Sigma_N(x_m)$ that can be efficiently computed by means of the Fast Fourier Transform algorithm, $I(\mathbf{x}) \approx FFT(\mathbf{x},\mathbf{h})$. The required values are computed with O$(N \log(N))$ operations. A thorough discussion on sampling and truncation errors is found in \cite{Lee04}.

\paragraph{Fractional Fourier Transform - FRFT.} The condition $\Delta \cdot \gamma = 2 \pi / N$ imposes that if we refine the integration grid ($\Delta$ small), the range for the variable $x$ becomes larger, thus including values which cannot be useful in our valuation procedure. Fractional Fourier transform permits on the contrary to decouple the two steps: it is therefore possible to choose properly the integration range and the $x$-spacing grid. The resulting algorithm, introduced in the financial literature in \cite{Cho05}, involves the use of standard FFT: in terms of the number of elementary operations, the computational cost of a FRFT procedure with $N$-point, $N$-FRFT, is about the same as a $4N$-FFT. The advantage of running a FRFT with smaller $N$ is that it may achieve the same accuracy than a FFT with much larger $N$.

The $m$-th component of the $\eta$-fractional discrete Fourier transform of the vector $\mathbf{h}$ is defined as
$$
FRFT(\mathbf{h},\eta)_m=\sum_{n=0}^{N-1} \e^{-\ci 2 \pi n m \eta} h_n, \ \ \
k=0,\ldots,N-1
$$
with $\eta = \Delta \gamma / 2 \pi$. The algorithm works as follows: firstly define two $2N$-point vectors
\begin{eqnarray*}
\mathbf{y} = (y_0, \ldots, y_{n-1}, y_{n}, \ldots, y_{2n}), &
y_j = h_j \e^{-\ci \pi j^2 \eta}, & 0 \leq j < n-1, y_j=0, \ \ n \leq j < 2n, \\
\mathbf{z} = (z_0, \ldots, z_{n-1}, \bar z_{n}, \ldots, \bar
z_{2n}), & z_j = \e^{\ci \pi j^2 \eta}, & 0 \leq j < n-1,  \bar
z_j = \e^{\ci \pi (n-j)^2 \eta}, \ 0 \leq j < n-1.
\end{eqnarray*}

The $m$-th component of $FRFT(\mathbf{h,\eta})_k$ is then computed as
$$
FRFT(\mathbf{h},\eta)_m = \e^{\ci \pi m^2 \eta} \odot
FFT^{-1}_m (FFT(\mathbf{y})_m \odot FFT(\mathbf{z})_m), \ \  m=1,
\ldots, n,
$$
where FFT$^{-1}$ is the inverse fast Fourier transform and $\odot$ is the component-wise vector multiplication. As before, the integral $I(x)$ can be approximated over the grid $\mathbf{x}$ by means of the Fractional Fourier Transform algorithm, $I(\mathbf{x}) \approx FRFT(\mathbf{x},\mathbf{h}, \eta)$

\section{Implementation and results}

The numerical procedures outlined in Section 2 for the computation of VaR and CVaR both require to evaluate the distribution function and/or the stop-loss expectation of the random variable $L$. Algorithm 1 firstly calls for a zero-finding routine that needs to compute (a sequence of) values of $F_L(\cdot)$ and then $SL(L)$ must be evaluated. Algorithm 2 must solve iteratively a univariate minimization problem requiring at each step to compute an expectation. Efficient numerical quadratures for computing Fourier integrals suffice for implementing the two algorithms without the need of calling for fast Fourier transforms. Error bounds on the approximation obtained clearly depends on the computational methods chosen for numerical integration, zero-finding and minimization routines. A third algorithm can be outlined, providing a quick-and-dirty solution, based on FFT/FRFT. Since we want to minimize w.r.t. $x$ the functions $G_{L,\alpha}(x)$ or $G^{(e)}_{L,\alpha}(x)$ we can approximate them for a whole range of values $\mathbf{x}$ by applying once FFT/FRFT: from the integral representation (\ref{Gfun0}) and (\ref{Gfun1}), we get
$$
G_{L,\alpha}(\mathbf{x})  \approx \hat G_{\alpha}(\mathbf{x}) \equiv \left \{ \begin{array}{l} \mathbf{x} - \frac{\e^{\nu \mathbf{x}}}{(1-\alpha) \pi} \odot FFT(\mathbf{x},\mathbf{h}) \\
\mathbf{x} - \frac{\e^{\nu \mathbf{x}}}{(1-\alpha) \pi} \odot FRFT(\mathbf{x},\mathbf{h},\eta)
\end{array} \right.
$$
and
$$
G^{(e)}_{L,\alpha}(\mathbf{x}) \approx \hat G^{(e)}_{\alpha}(\mathbf{x}) \equiv \left \{ \begin{array}{l} \mathbf{x} + \frac{\e^{\nu \mathbf{x}}}{(1-\alpha) \pi} \odot FFT(\mathbf{x},\mathbf{h}) \\
\mathbf{x} + \frac{\e^{\nu \mathbf{x}}}{(1-\alpha) \pi} \odot FRFT(\mathbf{x},\mathbf{h},\eta)
\end{array} \right.
$$
where $\mathbf{h}$ is the vector defined in (\ref{hvec}), evaluated according to the functions in (\ref{Gfun0}) or (\ref{Gfun1}). Hence, the minimum and the corresponding minimizer of the vector $\hat G_{L,\alpha}$ will provide approximated values for CVaR and VaR, respectively.

\begin{center}
\fbox{
\begin{minipage}{15cm}
\textsc{Algorithm 3. - \textsc{FFT/FRFT}}

\begin{enumerate}
        \item Compute the vector $\hat G_{L,\alpha}$ through FFT/FRFT algorithm for a proper grid $\mathbf{x}$;
        \item Find the minimum and the corresponding minimizer of $\hat G_{L,\alpha}$
\end{enumerate}
\end{minipage}
}
\end{center}

\bigskip

\begin{table}[h] \centering
\begin{tabular}{l|c|c|c} \hline
      & VaR Abs Err & CVaR Abs Err & Relative Time \\ \hline
\textsc{Alg 1} & $0.53 \times 10^{-14}$ & $0$ & $1$ \\
\textsc{Alg 2} & $0.33 \times 10^{-07}$ & $0$ & $0.3$\\
$2^{12}$-FFT    & $0.15 \times 10^{-02}$ & $0.32 \times 10^{-05} $ & $0.7 \times 10^{-05}$ \\
$2^{10}$-FRFT & $0.14 \times 10^{-03}$ & $0.27 \times 10^{-07}$ &
$0.2 \times 10^{-05}$\\ \hline \hline
\end{tabular}

\bigskip

\begin{tabular}{l|c|c|c} \hline
$n=5$, $p=0.1$   & VaR Abs Err & CVaR Abs Err & Relative Time \\
\hline
\textsc{Alg 2} & $0.0092$ & $0.0027$ & $1$\\
$2^{12}$-FFT    & $0.0106$ & $0.0025$ & $0.0092$ \\
$2^{10}$-FRFT & $0.0176$ & $0.0047$ & $0.0014$\\
\hline \hline
\end{tabular}

\caption{\small Comparison between the Fourier-based numerical procedures for the gaussian model (upper table), $X \sim N(\mu, \sigma^2)$ and the binomial model (lower table), $X \sim B(n,p)$, with $\alpha=0.99$. The second and third columns report absolute errors for the VaR and CVaR with respect to the true values, as obtained by applying zero-finding, univariate minimization (golden section search) and quadrature (adaptive Lobatto algorithm) build-in functions of MatLab. In view of the inversion results for discontinuous distribution functions, we did not apply Alg 1 to the binomial case. In these experiments, the zero-finding algorithm had starting point equal to mean of the r.v.; the univariate minimization requires a starting interval set to $[0,n]$ and $[0, \mu+ 3 \sigma]$, respectively. The FFT and FRFT algorithms have input vectors of length $2^N$; the integral was approximated between $0$ and $100$ in the gaussian case and $0$ and $200$ in the binomial case; the $x$-grid was started at $x_1=0$, with $\gamma=0.004$ for FRFT. In the last column the relative CPU times are shown, normalized to the slowest algorithm (Alg 1).} \label{tab1}
\end{table}

\subsection{Numerical results}

In this section we firstly report some results obtained by applying the computational procedures outlined above, that is Algorithm 1,2 and 3. We considered a gaussian and a binomial random variables for Table \ref{tab1}, and the univariate loss model framework of Example (\ref{CvarBS}) for Table \ref{tab2}: due to the analytical form of the function to be minimized (\ref{GfunEx}), we applied integral representation w.r.t. the scaled variable $x/V_0$. Pros and cons of each algorithm are easily outlined: precision and speed depend of course on the algorithms implemented, the programming language and on the computer available. In our experiment we used MatLab R2012a on a Intel Core i5 CPU with $2.40$ GHz. The basic steps of the algorithms (quadrature, univariate minimization, zero-finding and FFT) are those available as MatLab build-in functions.

\begin{table}[h] \centering
\begin{tabular}{l|c|c|c} \hline
$\mu=0$, $\sigma=0.2$, $T=\frac{1}{4}$ & VaR Abs Err & CVaR Abs Err & Relative Time \\ \hline
\textsc{Alg 1} & $0.11 \times 10^{-15}$ & $0.26 \times 10^{-14}$  & $1$ \\
\textsc{Alg 2} & $0.37 \times 10^{-08}$ & $0.22 \times 10^{-15}$  & $0.4628$\\
$2^{12}$-FFT   & $0.0011$                & $0.0017$               & $0.44 \times 10^{-03}$ \\
$2^{10}$-FRFT  & $0.14 \times 10^{-03}$  & $0.22 \times 10^{-05}$ & $0.58 \times 10^{-03}$\\
\hline \hline
\end{tabular}

\bigskip

\begin{tabular}{l|c|c|c} \hline
$\mu=-0.8$, $\sigma=0.35$, $T=\frac{1}{12}$ & VaR Abs Err & CVaR Abs Err & Relative Time \\ \hline
\textsc{Alg 1} & $0$                    & $0.55 \times 10^{-15}$ & $1$ \\
\textsc{Alg 2} & $0.36 \times 10^{-08}$ & $0$                    & $0.3583$\\
$2^{12}$-FFT   & $0.005$                & $0.0004$               & $0.0011$ \\
$2^{10}$-FRFT  & $0.88\times 10^{-04}$  & $0.23 \times 10^{-05}$  & $0.0013$\\
\hline \hline
\end{tabular}

\caption{\small Comparison between the numerical procedures for the benchmark univariate loss model - Example (\ref{CvarBS}) -  with $\alpha=0.99$. The second and third columns report absolute errors with respect to the true values for the VaR and CVaR. In these experiments, the zero-finding algorithm had starting point equal to the mid point of the interval $[0, V_0 \e^{rT}]$, while for  the univariate minimization the starting interval was set to $[0, \e^{rT}]$. The FFT and FRFT algorithms have input vectors of length $2^N$; the integrals were approximated between $0$ and $100$; the $x$-grid has right end point at $\log(V_0*\exp(r T))$, with $\gamma=6.7 \times 10^{-04}$ for FRFT. In the last column the relative CPU times are shown, normalized to the slowest algorithm (Alg 1).} \label{tab2}
\end{table}

\bigskip

In the second set of experiments, we show the effectiveness of the considered computational procedures in the univariate loss model by assuming different dynamics for $X_T$ and evaluating the impact of the relevant parameters on the computation of VaR and CVaR. The instances we consider are L\'{e}vy models with finite activity (Merton Jump-Diffusion) and infinite activity (Variance Gamma), and stochastic volatility models (Heston model) with jumps (Regime Switching Jump-Diffusion). But our procedure applies to all models characterized by having a computable (generalized) characteristic function. In such a cases, we used a hybrid approach consisting in two steps:
\begin{enumerate}
\item FRFT approximation for finding a feasible starting point $x_0$
\item refinement of the previous estimate by starting from $x_0$ a local minimization routine.
\end{enumerate}

In the following we simply report the GCF of the considered dynamics. Details can be found e.g. in \cite{CT}.

\paragraph{Merton Jump-Diffusion model.}  We consider a jump-diffusion setting in which the jump process is described as a marked point process (MPP). Let $\mu : \mathcal{S} \rightarrow \mathbb{R}$, $\sigma : \mathcal{S} \rightarrow \mathbb{R}$ and $\gamma : E \times \mathcal{S} \rightarrow \mathbb{R}$ be given functions, $(E,\mathcal{E})$ being the measurable mark space. Without loss of generality, we can assume in the following $E \subseteq \mathbb{R}$. In the given interval $0 \leq t \leq T$, we consider therefore the dynamic
\begin{equation}
X(t)  =  (\mu - \half \sigma^2) t + \sigma W(t) + \int_0^t \int_E \gamma(y)p(dy,ds),
\end{equation}
where $W(t)$ is a standard brownian motion and $p(dy,dt)$ is a MPP characterized by the intensity
$$
\lambda_t(dy) \equiv \lambda m(dy).
$$
Here $\lambda$ represents the intensity of the Poisson process $N_t$, while $m(dy)$ is a  probability measures on $E$ which specifies the jump variable $Y$. We assume that $W(\cdot)$ and $p(dy,dt)$ are independent and that $\E[\e^{\gamma(Y)}] = \int_E \e^{\gamma(y)} m(dy)$ is finite. The function $\gamma(y)$ represents the jump amplitude relative to the mark $y$: without loss of generality, we set $\gamma(y) =y$ in the following. The GCF for $X_T$ is then given by
$$
\phi_{X_T}(z) = \e^{\ci (\mu-\sigma^2/2) T z - \sigma^2 T z^2/2 + \lambda T (\phi_Y(z)-1)}
$$
where $\phi_Y(z) = \E[\e^{\ci z Y}]$. In the numerical example, we consider jumps characterized by a Normal distribution, $Y \sim N(a,b)$ so that
$$
\phi_Y(z) = \e^{\ci a z - b^2 z^2/2}.
$$

\begin{figure}
\begin{center}
\includegraphics[width=15cm,height=10cm]{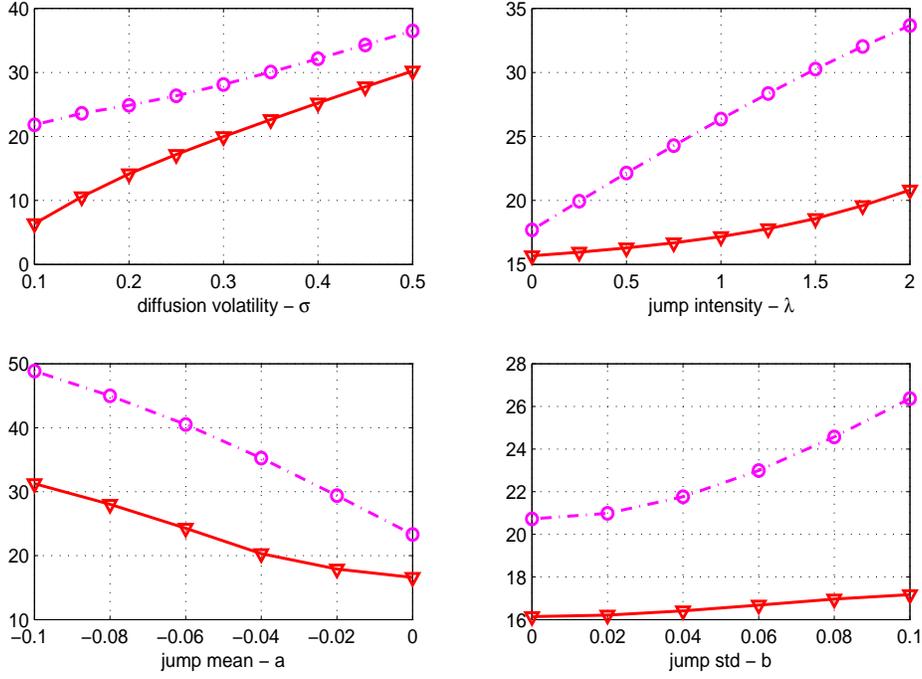}

\caption{\small  VaR (triangle) and CVaR (circle) for varying parameters of the \emph{Merton jump-diffusion model} with $V_0=100$, $r=0$, $T=1/12$, $\mu=0$, $\sigma=0.25$, $a=-0.01$, $b=0.1$ and
$\lambda=1$.} \label{FigMJD}
\end{center}
\end{figure}

VaR and CVaR obtained by varying the diffusion volatility $\sigma$, the jump intensity $\lambda$ and the jump parameters $a$ and $b$ are reported in Fig. (\ref{FigMJD}).

\paragraph{VG model.} The Variance Gamma model was introduced in \cite{MS90} and represents one of the simpler example of infinite activity L\`{e}vy model for describing an asset value dynamic. It can be defined as a Brownian
motion with drift, where time is changed by an independent gamma
process with mean rate unity and variance rate $\nu$, $G(t; 1,\nu)$:
$$
X_t = \theta  G(t; 1,\nu) + \sigma W_{G(t; 1,\nu)}.
$$

It has three parameters $\theta$, $\sigma$, $\nu$ and the characteristic function is given by
$$
\phi_{X_T}(z) = \left (\frac{1}{1- \ci \theta \nu z + \sigma^2 \nu
z ^2 /2} \right)^{T/\nu}.
$$

\begin{figure}
\begin{center}
\includegraphics[width=15cm,height=10cm]{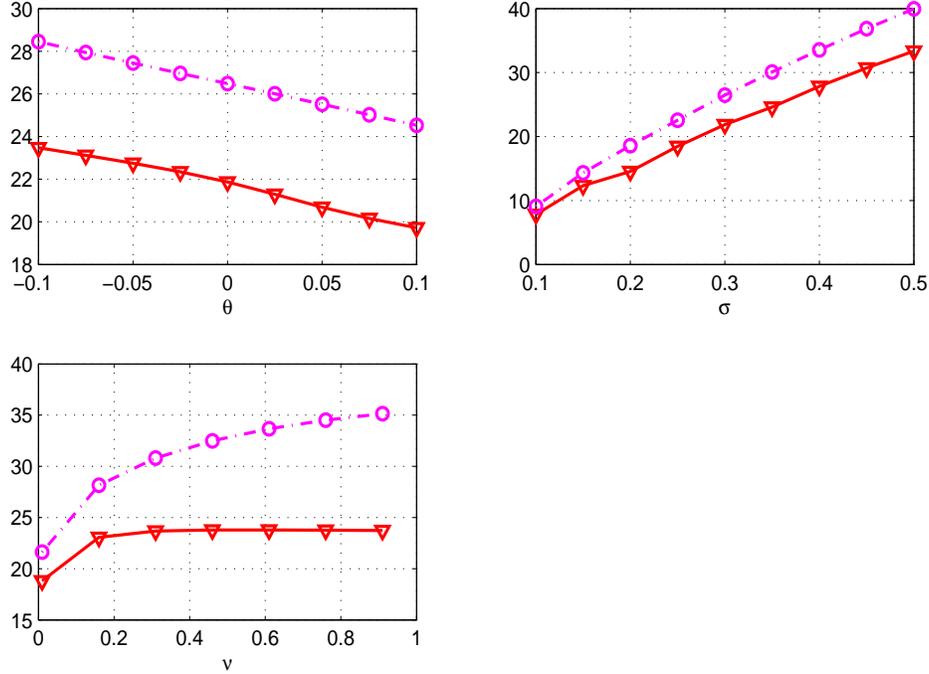}

\caption{\small  VaR (triangle) and CVaR (circle) for varying parameters of the \emph{Variance Gamma model} with $V_0=100$, $r=0$, $T=1/12$, $\theta=0$, $\sigma=0.3$, $\nu=0.1$.} \label{FigVG}
\end{center}
\end{figure}

In Fig. (\ref{FigVG}) the behavior of VaR and CVaR are compared for different values of the parameters.

\paragraph{Regime-Switching Jump-Diffusion model.} We consider a jump-diffusion model the parameters of which are driven by a finite state and continuous time Markov chain. To be definite, let $\omega(t)$ be a continuous time, homogeneous and stationary Markov chain on the state space $\mathcal{S} = \{1,2,\ldots,M \}$
with a generator $Q \in \mathbb{R}^{M\times M}$: the elements $q_{ij}$ of the matrix $Q$ are positive numbers such that $\sum_{j\neq i,j=1}^M q_{ij}=-q_{ii}$, for $i=1,\ldots,M$. The jump-diffusion dynamic is then modified as
\begin{equation} \label{rsjdm_sde}
X(t)  = \int_0^t (\mu(\omega(s)) - \half \sigma^2(\omega(s)) ds + \int_0^t \sigma(\omega(s)) d W(s) +  \int_0^t \int_E \gamma(y,\omega(t^-))p^{\omega}(dy,ds),
\end{equation}
where $p^{\omega}(dy,dt)$ is a MPP characterized by the regime-switching intensity $\lambda^{\omega}_t(dy) \equiv \lambda(\omega) m(\omega,dy)$, $m(\cdot,dy)$ being a set of probability measures on $E$, one for each state (regime) $i \in \mathcal{S}$. The function $\gamma(y,\omega)$ represents the jump amplitude relative to the mark $y$ in regime $\omega$. We assume that the processes $\omega(\cdot)$ and $W(\cdot)$ are independent, $W(\cdot)$ and $p^{\omega}(dy,dt)$ are conditionally independent given $\omega(t)$ and that $\E[\e^{\gamma(Y,\omega)}] = \int_E \e^{\gamma(y,\omega)} m(\omega,dy)$ is finite for each regime $\omega$.


In \cite{Ram10} (see also \cite{Cho05}) it was proved the following

{\proposition \label{propCFX} Let $\phi_j(z)=\E[\e^{\ci z \gamma(Y(j),j)}]$ be the generalized Fourier transform of the jump magnitude under the historical measure. Then, by letting
\begin{equation}\label{theta1}
\vartheta_j(z) = z (\mu(j) - \half \sigma^2(j)) + \half \ci z^2 \sigma^2(j) - \ci \lambda(j)(\phi_j(z)-1)
\end{equation}
and $\tilde \vartheta_i(z) = \vartheta_j(z) - \vartheta_M(z)$, we have
\begin{equation} \label{charfun2}
\begin{array}{lll}
\varphi_{X_T}(z) & = & \e^{\ci \vartheta_M(z) T} \left(\mathbf{1}' \cdot \e^{(Q' + \ci \ \mathrm{diag}(\tilde \vartheta_1(z), \ldots, \tilde \vartheta_{M-1}(z),0))T} \cdot \mathbb{I}(0) \right) \\ \\
 & = & \mathbf{1}' \cdot \e^{(Q' + \ci \ \mathrm{diag}(\vartheta_1(z), \ldots, \vartheta_{M}(z)))T} \cdot
\mathbb{I}(0), \end{array}
\end{equation}
where $\mathbf{1} = (1,\ldots,1)' \in \mathbb{R}^{M \times 1}$, $\mathbb{I}(0) = (\mathbb{I}_{\omega(0)=1}, \ldots, \mathbb{I}_{\omega(0)=M})'\in \mathbb{R}^{M \times 1}$ and $Q'$ is the transpose of $Q$. }

\bigskip

Simple linear constraints on the full parameter set of RSJD dynamic (\ref{rsjdm_sde}) permit to specify different models: from a regime-switching without jumps (the Naik model \cite{Nai} - RSGBM) to a unique regime jump-diffusion model (JDM), which includes the standard geometrical Brownian motion (GBM).

The evaluation of the characteristic function requires to compute matrix exponentials for which efficient numerical techniques are available (\cite{Higham09}); conversely, the case $M=2$ can be considered explicitly. The following can be proved (see \cite{Ram10} and the references therein).

{\proposition \label{twocharfun} Let $y_{1,2}$ be the solutions of the quadratic equation $y^2+(q_1+q_2-\ci \theta) y - \ci \theta q_2 = 0$ and
$$
\begin{array}{lll}
\mathrm{q}_1^{T}(\theta) & = & \frac{1}{y_1-y_2} \left( \e^{y_1 T}(y_1+q_1+q_2)-\e^{y_2 T} (y_2+q_1+q_2)\right) \\ \\
\mathrm{q}_2^{T}(\theta) & = & \frac{1}{y_1-y_2} \left(\e^{y_1 T}(y_1+q_1+q_2-\ci \theta)-\e^{y_2 T} (y_2+q_1+q_2-\ci \theta)
\right).
\end{array}
$$

Then
$$
\E_t[\e^{\ci \theta T_1}] = \mathbb{I}_{\omega(t)=1} \mathrm{q}_1^{T}(\theta)  +  \mathbb{I}_{\omega(t)=2}
\mathrm{q}_2^{T}(\theta)
$$
and therefore
$$
\varphi_{X_T}(z) = \e^{\ci \vartheta_2(z) T} \left( \mathbb{I}_{\omega(t)=1} \mathrm{q}_1^{T}(\theta(z)) +
\mathbb{I}_{\omega(t)=2} \mathrm{q}_2^{T}(\theta(z))  \right)
$$

$ \Box$ }

\begin{figure}[t]
\begin{center}
\includegraphics[width=15cm,height=10cm]{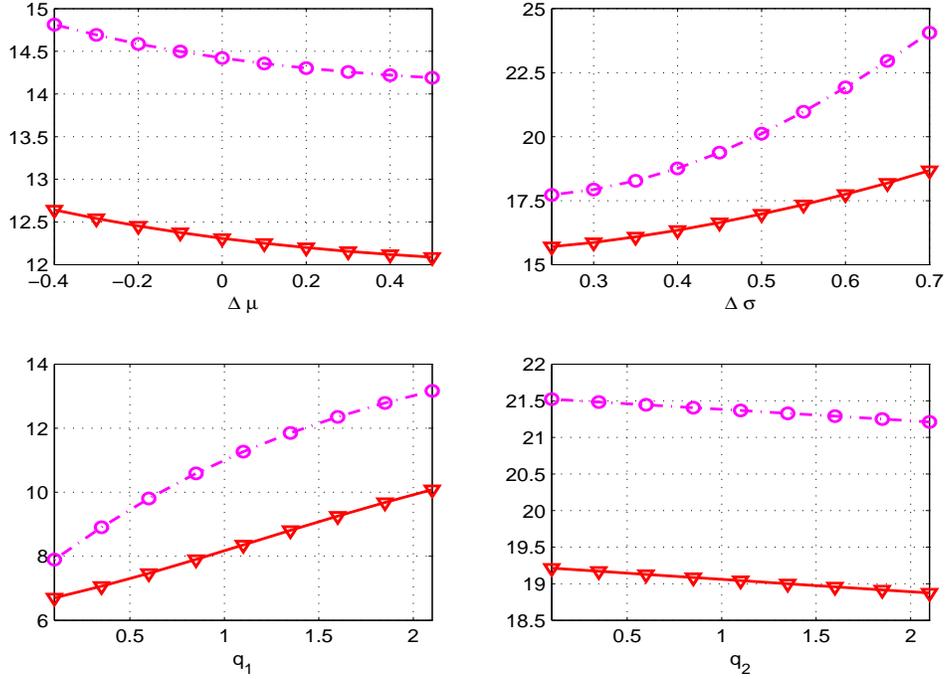}

\caption{\small  VaR (triangle) and CVaR (circle) for varying parameters of the \emph{RSGBM model} with $V_0=100$, $r=0$, $T=1/12$. In this model we consider a varying gap for the drift and volatility, $\mu_2=0.5+\Delta \mu$, $\sigma_2=0.1+\Delta \sigma$; when fixed the parameters were set to $\mu_1=0$, $\mu_2=-0.1$, $\sigma_1=0.1$, $\sigma_2=0.3$, $q_1=0.5$, $q_2=0.5$.} \label{FigRS}
\end{center}
\end{figure}

\begin{figure}[t]
\begin{center}
\includegraphics[width=15cm,height=10cm]{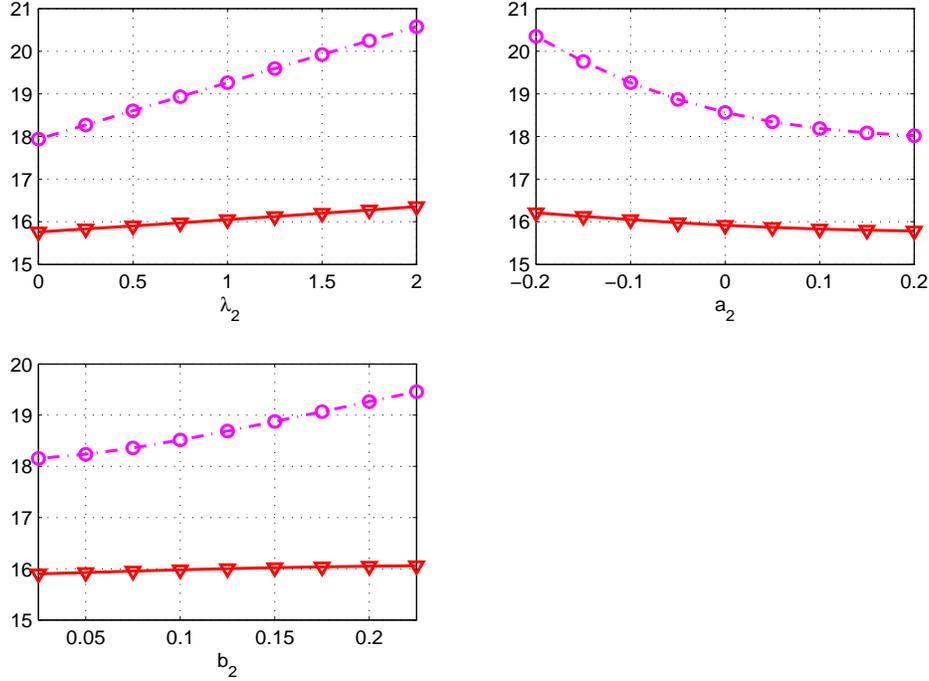}

\caption{\small  VaR (triangle) and CVaR (circle) for varying parameters of the \emph{RSJD model} with $V_0=100$, $r=0$, $T=1/12$. In this model we consider a fixed drift and volatility, $\mu=0$, $\sigma=0.25$ and vary the jump parameters with $\lambda_1=1$, $a_1=0.1$, $b_1=0.1$, and $q_1=0.5$, $q_2=0.5$.} \label{FigRSJD}
\end{center}
\end{figure}

Numerical tests are reported in Figs (\ref{FigRS}), (\ref{FigRSJD}) for a two-state model. In order to single out the effects of the switching parameters, we firstly consider the RSGMB model, thus discarding the jump component ($\lambda_1=\lambda_2=0$) - Fig (\ref{FigRS}); then we fix the diffusive dynamic ($\mu_1=\mu_2=\mu$, $\sigma_1=\sigma_2=\sigma$) and vary the jump parameters according to the switching model - Fig (\ref{FigRSJD}).

\paragraph{Heston Stochastic volatility model.} Heston model \cite{Hes} is certainly one of the most famous stochastic volatility dynamic for an asset price: it is defined as
\begin{eqnarray}
V_t & =& V_0 + \int_0^t V_s \mu ds + \int_0^t \sqrt{v_s} dW^1_s \\
v_t &=& v_0 + \int_0^t \kappa (\theta - v_s) ds + \sigma \int_0^t \sqrt{v_s}( \rho dW^1_s + \sqrt{1 -\rho^2}dW^2_s).
\end{eqnarray}
where $V_0 >0$, $\mu$ is the rate of return and $v_t$, the volatility process, satisfies a CIR mean reverting dynamic with parameters $\kappa$ (the mean reversion speed), $\theta$ (the long term volatility) and $\sigma$ (the vol-vol). Furthermore, the two process are $\rho$-correlated, with $-1 < \rho \leq 0$. The Feller condition $2 \kappa \theta > \sigma^2$ ensures the strict positivity of $v_t$. The (generalized) characteristic function of the log-price is
$$
\phi_{X_T}(z) = \e^{C(T,z) + D(T,z) v_0 + i z (\mu + \log(V_0))}
$$
where
\begin{eqnarray*}
C(T,z) & = & \frac{\kappa \theta}{\sigma^2} \left(
(\kappa - \rho \sigma z \ci + d(z))T - 2 \log \left( \frac{c(z) \e^{d(z)T}-1}{c(z)-1} \right)\right) \\
D(T,z) & = & \frac{\kappa -\rho \sigma z \ci + d(z)}{\sigma^2}
\left( \frac{\e^{d(z)T}-1}{c(z) \e^{d(z)T}-1} \right)
\end{eqnarray*}
and
$$
c(z) = \frac{\kappa - \rho \sigma z \ci + d(z)}{\kappa - \rho
\sigma z \ci - d(z)}, \ \ \ \ d(z) = \sqrt{(\rho \sigma z \ci-
\kappa)^2+\ci \sigma^2 z + \sigma^2 z^2}.
$$

\begin{figure}[t]
\begin{center}
\includegraphics[width=15cm,height=10cm]{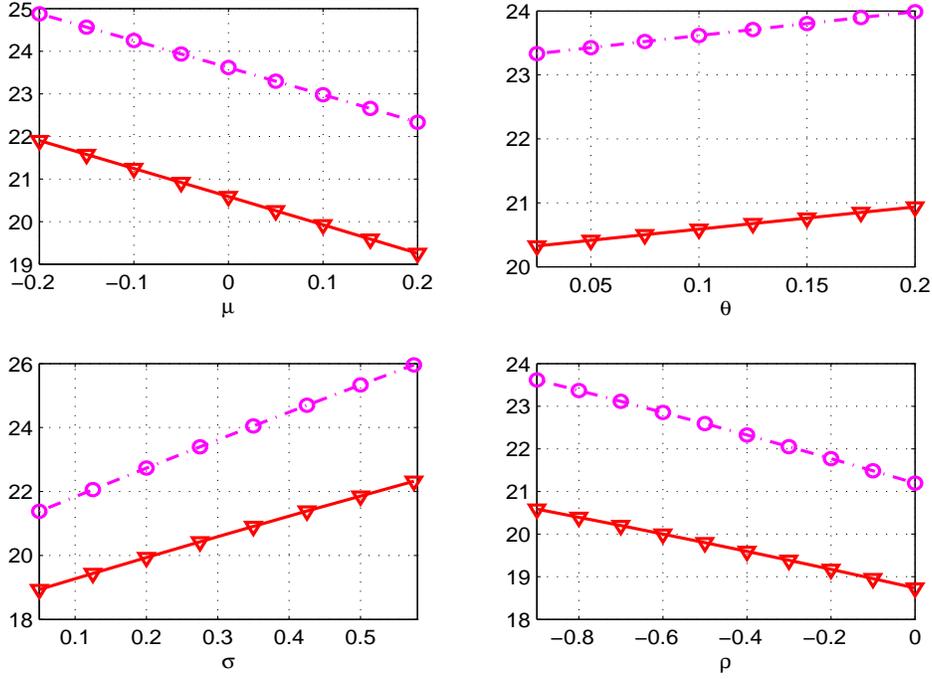}

\caption{\small  VaR (triangle) and CVaR (circle) for varying
parameters of the \emph{Heston model} with $V_0=100$, $r=0$,
$T=1/12$, $\mu=0$, $\theta=0.1$, $\sigma=0.3$, $\kappa=1$ and $\rho=-0.9$.} \label{FigHes}
\end{center}
\end{figure}

For the numerical implementation, we used the procedure outlined in \cite{KJ}. The corresponding results are plotted in Fig. (\ref{FigHes}).

\section{Conclusions}
In this paper we consider the problem of efficiently computing CVaR and VaR of an arbitrary loss function, characterized by having a computable generalized characteristic function. We compare different numerical procedures to compute the risk measures based on the integral representation of the distribution function and the stop/loss expectation of the target loss random variable. In particular we exploit the characterization of CVaR and VaR as solution of an univariate minimization problem, as obtained by Rockafellar and Uryasev in \cite{RU02}. The function to be minimized admits an integral representation as an inverse Fourier transform, under some hypothesis on the finiteness of the exponential moments of the loss distribution. Fast and reliable numerical procedures can be designed to compute the quantities of interest based on the fast Fourier transform algorithm. We finally notice that the basic characterization by Rockafellar and Uryasev is more general, since decision variables can be considered in the minimization problem: our procedure can therefore be included as part of more general optimization problem, like portfolio risk management, where the computation of VaR and CVaR plays a central role as objective functionals and/or constraints to be satisfied.

\end{document}